\documentclass[prl,amsmath,amssymb,twocolumn,showpacs]{revtex4}

\usepackage{amsmath,amssymb}
\usepackage[dvips]{graphicx}

\begin{document}

\title{Theory of RF-Spectroscopy of strongly interacting Fermions}

\author{M. Punk}
\author{W. Zwerger}
\affiliation{Physik-Department, Technische Universit\"at M\"unchen, James-Franck-Str., D-85748 Garching, Germany}

\date{\today}

\begin{abstract}
We show that strong pairing correlations in Fermi gases
lead to the appearance of a gap-like structure in the RF-spectrum,
both in the balanced superfluid and in the normal phase above the 
Clogston-Chandrasekhar limit. The average RF-shift of a unitary gas is proportional to the ratio of the Fermi velocity and the scattering length with the final state.
In the strongly imbalanced case, the RF-spectrum 
measures the binding energy of a minority atom to 
the Fermi sea of majority atoms. Our results provide a qualitative 
understanding of recent experiments by Schunck et.al. 
\end{abstract}


\maketitle


According to the fifty year old microscopic theory of 
Bardeen, Cooper and Schrieffer, the phenomenon of
superfluidity in a system of fermions is connected 
with the formation of bound pairs. In the weak coupling limit,
where the formation of pairs and their condensation
appears simultaneously, the transition to the superfluid state 
is associated with the appearance of a gap in the
fermionic excitation spectrum. For strong coupling, however,
this simple connection is no longer valid and bound pairs
of fermions may exist even in the normal state. 
This phenomenon is well known from the pseudogap phase in high temperature 
superconductors, where a d-wave pairing gap appears on the Fermi surface 
at temperatures far above the superconducting transition temperature \cite{Lee:2006}. 
A much simpler example is realized by ultracold fermions near a Feshbach resonance,
which provide a perfectly controllable model system to study the effects of strong 
pairing interactions \cite{Bloch}. In the case of an equal population of the two 
hyperfine states undergoing pairing, the ground state 
is superfluid at arbitrary values of the scattering length.
A microscopic signature of pairing in ultracold Fermi gases has first been 
obtained by Chin et.al. \cite{Chin} through RF-spectroscopy. 
The RF-field drives transitions between one of the hyperfine states 
$|2\rangle=|\!\!\downarrow\rangle$ which is involved in the pairing and an empty 
hyperfine state $|3\rangle$ which lies above it by an energy $\hbar\omega_{23}$ 
due to the magnetic field splitting of the bare atom hyperfine levels.
In the absence of any interactions, the spectrum 
exhibits a sharp peak at $\omega=\omega_{23}$. Pairing 
between the two lowest hyperfine states $|1\rangle$ and $|2\rangle$  
leads to an upward shift of this resonance. The shift essentially follows 
the two-particle binding energy on the BEC-side 
of the crossover but stays finite on the BCS-side,
where the appearance of a bound Cooper pairs
is a many-body effect \cite{Chin}.
A theoretical explanation of these observations can be given 
by extending the BCS description of pairing to the 
strong coupling regime and neglecting interactions 
involving state $|3\rangle$ \cite{Torma, Levin}. 
In a homogeneous system, the resulting RF-spectrum 
exhibits a peak at energies around $\Delta^2/\mu$, which is of 
the order of the energy gap $\Delta\approx 0.5\,\varepsilon_F$
at the unitarity point. 
Since pairing appears already in the normal state above $T_c$, the RF-shift does not directly measure the superfluid order, however \cite{Levin}. The importance of understanding the relation between RF-spectra and the nature of the many-body states involved, is underlined by recent experiments in imbalanced gases \cite{Schunck}.
There, a shift in the RF-spectrum 
is observed which hardly changes between the balanced superfluid and 
a normal ground state beyond a critical population imbalance,
where superfluidity is destroyed by a sufficiently large 
mismatch of the Fermi energies even at $T=0\/$ (this is the analog of the Clogston-Chandrasekhar limit in superconductors).
In this work, we present a theory of RF-shifts in both balanced
and imbalanced Fermi gases, which 
provides a qualitative understanding of these observations. 
In particular, we show that the average frequency shift in the balanced 
superfluid at unitarity (i.e. at infinite scattering length) is linear in the Fermi velocity and 
inversely proportional to the scattering length $a_{13}$.  
In the non-superfluid state beyond the Clogston-Chandrasekhar limit,
pair fluctuations give rise to sharp peaks in the RF-spectrum which are
associated with the binding of $\uparrow\,\downarrow$-pairs even
in the absence of long range phase coherence.

Within linear response theory, which is adequate for RF pulses 
short compared to the Rabi oscillation period of the bare 2-3 transition, 
the number of particles transferred
from state $|2\rangle$ to state $|3\rangle$ per unit time is given by
\begin{eqnarray}
I(\omega) \! &\sim& \! \int dt \; d^3\!x \; d^3\!x' e^{i (\mu_3-\mu_\downarrow-\omega_L) t} \notag \\
\! &\times& \! \left\langle \left[ \psi^\dagger_3(x,t) \psi_\downarrow(x,t),\psi^\dagger_\downarrow(x',0) \psi_3(x',0)\right] \right\rangle 
\label{equ:spectralcurrent}
\end{eqnarray}
where $\omega=\omega_L-\omega_{23}$ denotes the detuning of the RF field from the bare 2-3 transition. Since particles in state $|3\rangle$ have a nonvanishing interaction with those in states $|1\rangle$ and $|2\rangle$ \cite{Gupta}, the response function in equation (1) does not factorize into one particle functions, making a full calculation of the spectrum very difficult.
Nevertheless, near $T=0$, where only a single peak is observed in the RF-spectrum, its position can be determined from a sum rule approach \cite{Yu}. In particular, the first moment 
$\bar{\omega}=\int d\omega\,\omega I(\omega) / \int d\omega\, I(\omega)$ is given by
\begin{equation}
\hbar\bar{\omega}=\frac{\bar{g}_{12}-\bar{g}_{13}}{N_2-N_3} \left( \frac{\langle H^{'}_{13}\rangle}{\bar{g}_{13}}-\frac{\langle H^{'}_{12}\rangle}{\bar{g}_{12}} \right)\, .
\label{equ:sumrule1}
\end{equation}
Here $H^{'}_{13}$ and $H^{'}_{12}$ denote the interaction Hamiltonians between the respective states,  while $N_2$ and $N_3$ denote the total number of particles in states $|2\rangle$ and $|3\rangle$. 
The $\bar{g}_{ij}$ are the bare interaction constants arising in the pseudopotential interaction
Hamiltonian
\begin{equation}
H^{'}_{ij}=\bar{g}_{ij} \int d^3x \; \psi^\dagger_i(x)\psi^\dagger_j(x) \psi_j(x) \psi_i(x)\, .
\label{equ:hamiltonian}
\end{equation}
They are related to their renormalized values  
$g_{ij}= 4 \pi \hbar^2 a_{ij}/m$ by
\begin{equation}
\frac{1}{\bar{g}}=\frac{1}{g}-\int \frac{d^3k}{(2 \pi)^3} \; \frac{1}{2 \varepsilon_\mathbf{k}}
\end{equation}
where $a_{ij}$ are the s-wave scattering lengths between states $i$ and $j$, $m$ is the mass of the particles and $\varepsilon_\mathbf{k}=\hbar^2 k^2/2 m$ the single particle energy.
Note that the interaction $g_{23}$ between states 2 and 3 drops out quite generally, because $H^{'}_{23}$ and $H_\text{RF}$ commute. Moreover, there is no shift of the RF peak if the interaction strengths $g_{12}$ and $g_{13}$ are equal, a case, where all interaction effects 
are cancelled exactly \cite{Yu,Zwierlein}. Since $\langle H^{'}_{13}\rangle$ is of order $N_3$, the first term in (\ref{equ:sumrule1}) is negligible compared to the second term if $N_2 \gg N_3$. 
The average shift of the RF-spectrum then simplifies to
\begin{equation}
\hbar\bar{\omega}=\frac{\langle H^{'}_{12}\rangle}{N_2}
\left(\frac{\bar{g}_{13}}{\bar{g}_{12}}-1 \right)\to 
\frac{\langle H^{'}_{12}\rangle}{N_2\Lambda}\frac{\pi}{2}
\left(\frac{1}{a_{13}}-\frac{1}{a_{12}} \right)\, .
\label{equ:sumrule2}
\end{equation}
Here, the second form is obtained by expanding $1-\bar{g}_{13}/\bar{g}_{12}$
to leading order in the upper cutoff $\Lambda$ of the momentum integral
in (4). Evidently, for vanishing interactions $\bar{g}_{13}=g_{13}\equiv 0$ with state 3,
the RF-shift just measures the (negative) interaction energy per particle in the 
state 2. Within a pseudopotential description, however, the interaction 
energy $\langle H^{'}_{12}\rangle\sim\Lambda$ diverges linearly with
the cutoff.  It is thus sensitive to the range of the interactions, which is set equal to zero 
in the pseudopotential. In terms of the spectrum $I(\omega)$, this divergence 
shows up as a slow decay $I(\omega)\sim\omega^{-3/2}$ at large frequencies,
leading to a divergent first moment, as is easily seen within a 
BCS-description with a constant gap $\Delta$. Remarkably, for finite interactions 
$g_{13}\ne 0$, the second form of (5) gives a result for the frequency shift which is
well defined and finite in the limit $\Lambda\to\infty$. As shown by Tan \cite{Tan},
the total energy of the balanced gas can be obtained from the momentum 
distribution $n_{\mathbf{k}}$ via 
$E=2\sum_{k}\varepsilon_{\mathbf{k}}(n_{\mathbf{k}}-C/k^4)$ up to a constant,
which is irrelevant for the calculation of the limit $\langle H^{'}_{12}\rangle/\Lambda$.
Here $C$ is the constant arising in the asymptotic behavior $\lim n_{\mathbf{k}}
=C/k^4$ of the momentum distribution at large momenta.
Evidently, the interaction contribution to the total energy 
is just  $\langle H^{'}_{12}\rangle=-2C\sum_k\varepsilon_{\mathbf{k}}/k^4\sim-C\Lambda$. Introducing 
a dimensionless constant $s$ via $C=sk_F^4$, the shift of the RF-spectrum
\begin{equation}
\hbar\bar{\omega}=s\cdot\frac{4\varepsilon_F^2}{n_2}
\left(\frac{1}{g_{12}}-\frac{1}{g_{13}}\right)
\end{equation}
of the balanced gas is completely determined by the universal constant $s$, the 
Fermi energy $\varepsilon_F= \hbar^2 k_F^2/(2 m)$ of the balanced, non-interacting gas and the renormalized interaction constants $g_{12}$ 
and $g_{13}$. The expression is finite for all coupling strengths $g_{12}$
and evolves smoothly from the BCS- to the BEC-limit. 
Within an extended BCS-description of the ground state wavefunction, the product $s^{(0)}\cdot 4\varepsilon_F^2\equiv\Delta^2$ is precisely the square of the gap parameter. In weak coupling, our result then coincides with that obtained by Yu and Baym \cite{Yu}, except for the mean field shift, which is not contained in the reduced BCS Hamiltonian.
In the BEC-limit, where the BCS-groundstate becomes exact, the asymptotic behavior 
$\Delta_{BEC}=4\varepsilon_F/\sqrt{3\pi k_Fa_{12}}$ gives $\hbar\bar{\omega}=2\varepsilon_b (1-a_{12}/a_{13})$,
where $\varepsilon_b=\hbar^2/ma_{12}^2$ is the two-particle binding energy.  
It is straightforward to show, that this is precisely the average shift for bound-free transitions following from a detailed calculation of the RF-spectrum in the molecular limit by Chin and Julienne \cite{ChinJulienne}.
The most interesting regime is that around
the unitarity limit $1/g_{12}=0$. At this point, the average RF-shift is given by 
$\bar{\omega}=-0.46\,v_F/a_{13}$, which varies like the \emph{square root} of the Fermi energy $\varepsilon_F=m v_F^2/2$. The constant
$s= 0.098$ is obtained  from the recent calculations of
the crossover thermodynamics by Haussmann et al. \cite{Haussmann}.
Our result for the homogeneous gas can be compared directly with locally resolved RF-spectra by Shin et al. \cite{Shin}.
Accounting for the enhancement of the local Fermi velocity at the trap center by a factor $\approx 1.25$ due to the attractive interactions, the predicted average shift $\bar{\omega}=2 \pi \cdot 28.9 \text{kHz}$ \cite{Bartenstein:2004} is considerably larger than the measured position of the peak near 15 kHz. This is probably due to the fact, that $\bar{\omega}$ has a considerable contribution from the higher frequency part of the spectrum. A crucial prediction of our theory is the linear behaviour of the average RF-shift with the Fermi momentum. Experimentally, the spatial resolution necessary to distinguish this from the naive $\varepsilon_F$-scaling has not yet been achieved \cite{Shin}.

\begin{figure}
	\begin{center}
		\includegraphics[width=0.9\columnwidth]{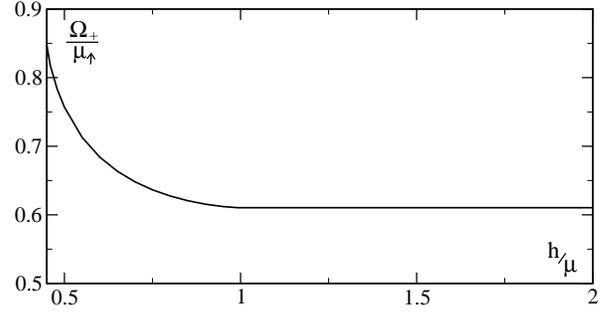}
	\end{center}
	\caption{Pair binding energy $\Omega_+$ in units of $\mu_\uparrow$ at unitarity as a function of $h$ at T=0. For $h>\mu$ the binding energy is constant and given by $\Omega_+=0.61 \mu_\uparrow$.}
	\label{fig:bind_en}
\end{figure}

To discuss the situation with a finite imbalance, 
it is convenient to introduce two distinct chemical potentials 
for the states undergoing pairing, defined by $\mu_\uparrow=\mu+h$ and $\mu_\downarrow=\mu-h$.
Since the ground state of the spin balanced 
gas is a superfluid with a gap for fermionic excitations, it will 
be stable over a finite range $h<h_c$ of the chemical potential difference.
In the BCS limit, the associated Clogston-Chandrasekhar 
critical field $h_{\rm c}=\Delta_{\rm BCS}/\sqrt{2}$ is exponentially small.
Near the unitarity point,  the absence of a second energy scale 
implies that the critical field $h_c$ beyond which a non-zero 
polarization appears, is on the order of the bare Fermi energy
$\varepsilon_F$ of the balanced two-component Fermi gas. 
From fixed node diffusion Monte Carlo calculations 
the resulting numerical value in the continuum case at unitarity is 
$h_c=0.96 \,\mu\approx 0.4\,\varepsilon_F$ \cite{Lobo:2006}. 
The phase for $h>h_c$ is a non-superfluid, polarized 
mixture of the different spin states. 
For large enough fields, the system will eventually be 
completely spin polarized.  At unitarity, the associated saturation field
$h_s$  was determined by Chevy \cite{Chevy} using
a variational calculation of the energy change $\mu_{\downarrow}$
associated with adding a single $\downarrow$-particle to 
a Fermi sea of $\uparrow$-particles.  This  
leads to an upper bound $\mu_{\downarrow}\leq -0.60\,\mu_{\uparrow}$ 
at the unitarity point,
where $\mu_{\uparrow}=2^{2/3}\varepsilon_F$ is the Fermi 
energy of the completely spin polarized gas.  The 
saturation field thus obeys the inequality $h_s\geq 0.8\,\mu_{\uparrow}=1.27\,\varepsilon_F$.
At unitarity, therefore, there is a wide regime $h_c<h<h_s$ 
of an intermediate phase between the balanced superfluid and 
a fully polarized gas. While superfluidity is 
quenched in this phase, the strong interactions between 
particles in states $|1\rangle$ and $|2\rangle$ still give rise to large frequency shifts
in the RF-spectrum, as will be shown below.
To study the effect of pairing fluctuations on the imbalanced Fermi gas above the Clogston-Chandrasekhar limit, we calculate the pair-fluctuation spectrum from the two-fermion Green function, using a non-selfconsistent T-matrix approach, similar 
to the approach by Combescot et al. \cite{Combescot:2007}. Such a perturbative analysis is reasonable, since the states which are coupled through the interaction Hamiltonian are separated by an energy gap of width $2 h$. A usual ladder approximation is used to incorporate the effects of the attractive $\uparrow \, \downarrow$-interaction on the vertex part, whereas the self energy is calculated at the one-loop level, including vertex corrections. The basic equations for the polarization loop $L$, vertex part $\Gamma$ and self-energy part of the minority species $\Sigma_\downarrow$ are given by (we take units in which $\hbar=1$)
\begin{figure}
	\begin{center}
		\includegraphics[width=0.9\columnwidth]{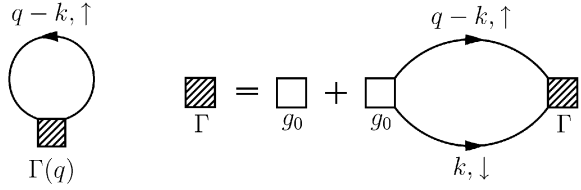}
	\end{center}
	\caption{Lowest order self energy diagram for the minority component $(\downarrow)$ Green's function and Bethe-Salpeter equation for the vertex part in ladder approximation.}
\end{figure}
\begin{eqnarray}
L(\mathbf{q}, i \Omega_n) &=& -\frac{1}{\beta} \sum_{\omega_m} \int \frac{d^3 k}{(2 \pi)^3} \mathcal{G}^{(0)}_\uparrow (\mathbf{q}-\mathbf{k},i\Omega_n-i\omega_m) \notag \\
&\times& \mathcal{G}^{(0)}_\downarrow (\mathbf{k},i\omega_m) \\
\Gamma(\mathbf{q},i\Omega_n) &=& \frac{1}{1/\bar{g}-L(\mathbf{q},i\Omega_n)} 
\end{eqnarray}
\begin{eqnarray}
\Sigma_\downarrow(\mathbf{k},i \omega_n) &=& \frac{1}{\beta} \sum_{\Omega_m} \int \frac{d^3 q}{(2 \pi)^3} \Gamma(\mathbf{q},i \Omega_m) \notag \\
&\times& \mathcal{G}^{(0)}_\uparrow(\mathbf{q}-\mathbf{k},i\Omega_m-i\omega_n) \label{matsub_self_en}
\end{eqnarray}
where $\mathcal{G}^{(0)}_\uparrow$ and $\mathcal{G}^{(0)}_\downarrow$ are the bare Matsubara-Green's functions of the majority- and minority component and $\beta=1/k_B T$ is the inverse temperature. $\omega_n=(2n+1)\pi/\beta$ and $\Omega_n=2\pi n/\beta$ with $n \in \mathbb{Z}$ denote fermionic and bosonic Matsubara frequencies respectively.
After evaluating the Matsubara summation and analytic continuation, the vertex part can be calculated analytically at $T=0$. In the regime $h>\mu$ (i.e. essentially beyond the Clogston-Chandrasekhar field $h>h_c\approx 0.96 \mu$), one obtains for $\mathbf{q}=0$, $\omega>-2 \mu$
\begin{eqnarray}
\Gamma^R(\mathbf{0},\omega) \! &=& \! \frac{2 \pi^2}{m k_{F \uparrow}} \Bigg\{-\frac{\pi}{2 k_{F \uparrow} |a|} - 1 +  \frac{1}{2} \sqrt{\frac{\omega+2 \mu}{2 \mu_\uparrow}} \notag \\
&\times& \! \Bigg[ \ln \left| \frac{1+\sqrt{\frac{\omega+2 \mu}{2 \mu_\uparrow}}}{1-\sqrt{\frac{\omega+2 \mu}{2 \mu_\uparrow}}} \right| + \text{i} \pi \, \Theta(\omega-2 h) \Bigg]  \Bigg\}^{-1}
\label{equ:vertex}
\end{eqnarray}
where $\Theta(x)$ is the unit step function and $k_{F \uparrow}$ is defined via $k_{F \uparrow}=\sqrt{2 m \mu_\uparrow}/\hbar$.
For $h>\mu$ the retarded vertex $\Gamma^R(\mathbf{q}=0,\omega)$ has a single pole on the real axis at
$\omega_0^+ = 2 h - \Omega_+$ with $\Omega_+ >0$ (note that for $h <\mu$ the vertex has two real poles).
Physically, this pole describes an excitation in which two fermions with opposite spin and vanishing total momentum form a pair at the Fermi energy of the majority component with binding energy $\Omega_+$.
A similar structure was first discussed for weak coupling by Aleiner and Altshuler \cite{Aleiner} in the context of small superconducting grains.  Remarkably, as shown in Fig. \ref{fig:bind_en}, the pair binding energy in units of $\mu_\uparrow$ is constant for $h>\mu$ and agrees well with the value $0.6 \mu_\uparrow$ for the binding energy of a single down spin in the presence of a Fermi sea of majority atoms as calculated by Chevy \cite{Chevy}.
The retarded self energy for the minority component in the normal state is given by
\begin{eqnarray}
\Sigma_{\downarrow}^R(\mathbf{k},\omega) \! &=& \! \int \frac{d^3q}{(2\pi)^3} \, \frac{dz}{\pi} \, \Big\{ n_B(z) \, G_{A,\uparrow}^{(0)}(\mathbf{q-k},z-\omega) \notag \\
&\times& \text{Im} \Gamma^R(\mathbf{q},z)-n_F(z) \, \text{Im} G_{R,\uparrow}^{(0)}(\mathbf{q-k},z) \notag \\
&\times& \Gamma^R(\mathbf{q},z+\omega) \Big\}
\end{eqnarray}
with $n_B$ and $n_F$ denoting the Bose- and Fermi-distributions.
\begin{figure}
	\vspace*{0.5cm}
	\begin{center}
		\includegraphics[width=0.8\columnwidth]{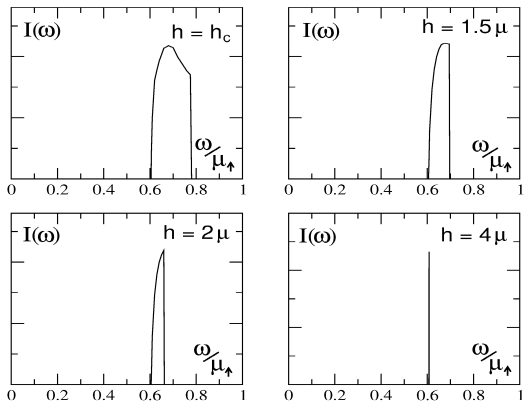}
	\end{center}
	\caption{RF spectra at unitarity for different imbalances at $T=0$ (Intensity in arbitrary units; $h_{c}=0.96 \mu$).}
	\label{fig:spektren}
\end{figure}
This result enables us to calculate RF spectra explicitly in the limit of vanishing $g_{23}$ and $g_{13}$, where the expactation value in equ. (\ref{equ:spectralcurrent}) can be factorized. In this case,
one obtains \cite{Torma,Levin}
\begin{eqnarray}
I(\omega) \! &\sim& \! \int \! \frac{d^3k}{(2\pi)^3} \; \text{Im} G_{\downarrow}^R(\mathbf{k},\varepsilon_{\mathbf{k}}-\omega-\mu_\downarrow) \; n_F(\varepsilon_{\mathbf{k}}-\omega-\mu_\downarrow) \notag
\end{eqnarray}
if state $|3\rangle$ is initially empty. In figure \ref{fig:spektren} we have numerically evaluated the resulting RF-spectra at unitarity for different fields above the Clogston-Chandrasekhar limit.
The calculation explains two features which are seen in the experimental data \cite{Schunck}, namely the shift of the RF peak due to pairing fluctuations in the normal state and the decreasing linewidth with increasing population imbalance. The onset of the RF-spectrum coincides with the pair binding energy
$\Omega_+\approx 0.6\,\mu_{\uparrow}$ for $h>h_c$, which is independent of the imbalance.
In the presence of a finite $|1\rangle-|3\rangle$ interaction, the detailed spectrum $I(\omega)$ can not be calculated analytically. Its first moment, however, is again determined by the sumrule equ. (\ref{equ:sumrule2}). Evaluating the interaction energy $\langle H^{'}_{12}\rangle$ using the variational wavefunction of Chevy \cite{Chevy}, it turns out that the resulting average RF-shift for an almost completely polarized gas is equal to $\bar{\omega}=-0.34\,\hbar k_{F\uparrow}/ma_{13}$. Due to the sharpness of the peak in this limit, the average shift in the strongly imbalanced gas coincides with the experimentally observed peak position. For the parameters in \cite{Shin}, we obtain an average RF-shift $\bar{\omega}=2 \pi \cdot 17 \, \text{kHz}$ at the trap center for strong imbalance, close to the observed value in the balanced case. Our theory thus accounts for the observation by Schunck et al. \cite{Schunck}, where an average over the trap is involved, that there is hardly any difference in the RF-shift between the balanced and strongly imbalanced gas.

In conclusion, we have given a theory of RF-spectra in ultracold Fermi gases which 
includes interactions between all three states involved.  In the balanced
unitary gas, the average RF-shift is proportional to $-s\, v_F/a_{13}$, where $s$ is a universal constant characterizing the fermion momentum distribution 
at large wave vectors. In the imbalanced case, the RF-spectrum exhibits a sharp peak arising from the binding
energy of a $\uparrow\,\downarrow$-pair which is finite even in the non-superfluid state. Including a finite value of $a_{13}$, the resulting average shift is close to the peak shift in the balanced case. 

We gratefully acknowledge very helpful discussions with W. Rantner, Yong-il Shin and M. Zwierlein. This work was supported 
by the DFG Forschergruppe "Strong Correlations in multiflavor ultracold Quantum Gases".

Note added in proof: Equivalent results for the RF-shift of balanced gases have been obtained independently by Baym et al. \cite{Baym}. In fact our value for the prefactor in $\bar{\omega}=-0.46\,v_F/a_{13}$ agrees well with the value obtained in this reference, using a different method.

\end{document}